\documentclass[pof,aps,preprint,a4paper]{revtex4} 
\usepackage{graphicx}
\usepackage{color}

\graphicspath{{PLOTS/}}

\begin{document}

\title{Hydrodynamics of air entrainment by
    moving contact lines}

\author{T.S. Chan$^1$, S. Srivastava$^{2,5}$, A. Marchand$^3$,
  B. Andreotti$^3$, L. Biferale$^4$, F. Toschi$^{2,5,6}$ and
  J.H. Snoeijer$^1$} \affiliation{ $^1$ Physics of Fluids Group,
  Faculty of Science and Technology and MESA+ Institute, University of
  Twente, 7500AE Enschede, The Netherlands\\ $^2$ Department of 
Applied Physics, Eindhoven University of Technology, 5600 MB Eindhoven, The
  Netherlands\\ $^3$ Physique et M\'ecanique des Milieux
  H\'et\'erog\`enes, UMR 7636 ESPCI -CNRS, Universit\'e Paris-Diderot,
  10 rue Vauquelin, 75005, Paris\\ $^4$ Department of Physics and
  INFN, University of Tor Vergata, Via della Ricerca Scientifica 1,
  00133 Rome, Italy\\ $^5$ Department of Mathematics and Computer
  Science Eindhoven University of Technology, 5600 MB Eindhoven The
  Netherlands \\ $^6$ CNR-IAC, Via dei Taurini 19, 00185 Rome, Italy }

\date{\today}

\begin{abstract}
We study the dynamics of the interface between two immiscible fluids
in contact with a chemically homogeneous moving solid plate.  We
consider the generic case of two fluids with any viscosity ratio and
of a plate moving in either directions (pulled or pushed in the
bath). The problem is studied by a combination of two models, namely
an extension to finite viscosity ratio of the lubrication
theory and a Lattice Boltzmann method.  Both methods allow to resolve,
in different ways, the viscous singularity at the triple contact
between the two fluids and the wall.  We find a good agreement between
the two models particularly for small capillary numbers.  When the
solid plate moves fast enough, the entrainment of one fluid into the
other one can occur.  The extension of the lubrication model to the
case of a non-zero air viscosity, as developed here, allows us to
study the dependence of the critical capillary number for air
entrainment on the other parameters in the problem (contact angle and
viscosity ratio).
\end{abstract}

\maketitle

\section{Introduction}
The problem of a solid plate pulled from a liquid bath has attracted
considerable attention in the past including the seminal contributions
from Landau, Levich, Derjaguin and Bretherton \cite{LL42,D43,B61}. 
The problem has continued to be investigated, with a particular 
focus on the situation of partial wetting for which a dynamical wetting 
transition is observed: a liquid film is deposited only when a critical speed 
of withdrawal is exceeded (see \cite{BEIMR09,Snoeijer13} for reviews). 
Much less is known about the reversed case, where the solid is plunged into the liquid. 
Again, a dynamical wetting transition has been observed, now resulting 
into the entrainment of an air film or air bubbles \cite{Ben07,Ben10,MCSA12, DYCB07, LRHP11,BK76a}. 
Despite the viscosity contrast between the liquid in the reservoir and that 
of the surrounding air, the dynamics inside the air is very important for this 
process. The perturbation analysis by Cox~\cite{C86} suggested that the
critical speed is inversely proportional to the
  viscosity of the liquid, $~1/\eta_\ell$, with logarithmic
corrections due to the viscosity of the air, $\eta_g$. This is similar
to the scenario for air entrainment by viscous cusps~\cite{E01}, such
as observed for impacting liquid jets \cite{LRQ03}. 

Recently, in the
experiment of plunging a plate into reservoirs of different liquids
 Marchand \emph{et al.} \cite{MCSA12}
observed that the dependence on $\eta_\ell$ is much weaker than
predicted; enforcing a power-law fit to their data would give a small
exponent, in between -1/2 and -1/3 rather than the expected -1. This
implies that air viscosity plays an important role on the onset of air
entrainment even if it is orders of magnitude smaller than the liquid
viscosity. The importance of air was already highlighted in similar
dip-coating experiments, where a reduction of the ambient pressure was
shown to significantly enhance the critical speed of entrainment
\cite{Ben07}. Yet, this raises another paradox: the dynamical
viscosity of the air is virtually unchanged by a pressure reduction.

In this paper we provide a new framework to study air entrainment by
advancing contact lines, in which the two-phase character of the flow
is taken into account. The usual lubrication approximation is valid
for small (liquid) contact angles and does not take into account the
flow inside the gas. Since both assumptions are no longer valid near
the onset of air entrainment, we extend the lubrication approximation
such that it allows for large angles and a nonzero gas viscosity. This
generalized lubrication theory is validated by comparing to Lattice Boltzmann
simulations. We show that the meniscus shapes obtained from the
generalized lubrication model agrees well with the simulations. Note,
however, that such simulations are limited in terms of viscosity ratio
and spatial resolution (i.e. the separation between the capillary
length and the microscopic cutoff given by the interface width or,
equivalently, by the effective slip length \cite{Sbragaglia2008}), and
cannot achieve experimental conditions for air entrainment. To compare
to experiments and to explore the parameter space, in particular the
importance of air viscosity $\eta_g$, we therefore provide a detailed
study using the generalized lubrication model.

\section{Formulation}

\begin{figure}
\begin{center}
\includegraphics[width=0.5\textwidth]{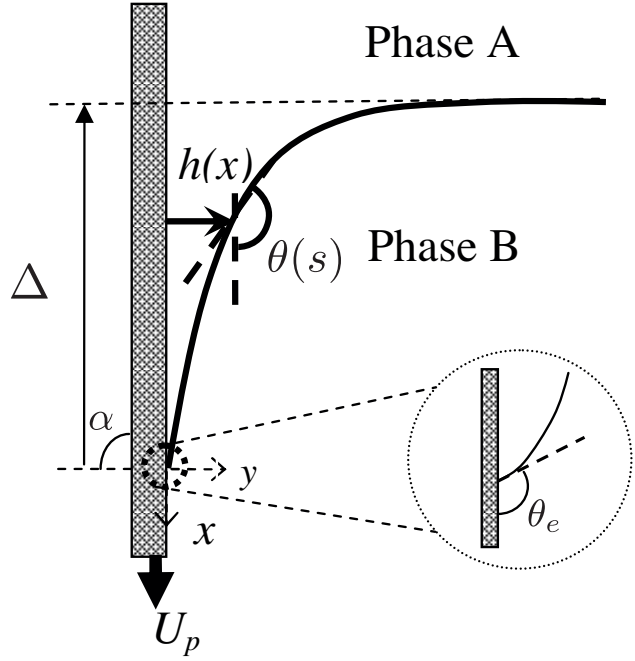}
\caption{\label{fig1setup}Schematic diagram showing a plate moving
  with speed $U_p$ with respect to an interface of two immiscible
  fluids. The plate has an inclination $\alpha$, here drawn with
  $\alpha=\pi/2$ for the case of a vertical plate. The interface
  touches the wall at the contact line with an angle assumed to be the
  same as the equilibrium contact angle $\theta_e$. The meniscus
  profile is described by $h(x)$ or the local angle $\theta(s)$, where
  $s$ is the arc length of the interface measured from the contact
  line. The total meniscus deformation is quantified by $\Delta$, the
  distance between the contact line and the level of the bath.}
\end{center}
\end{figure}

We consider a smooth, chemically homogeneous solid plate translating
across an interface of two immiscible fluids at a constant speed $U_p$
(positive/negative for plunging/withdrawing). As sketched in
Fig. \ref{fig1setup}, the two fluids are contained in a reservoir
much larger than all the lengths of the problem and the plate can be
inclined to any angle $\alpha$. If the plate is not moving ($U_p=0$)
there is no flow in the fluids, and the interface equilibrates to a
static shape due to balance between capillarity and gravity. The
interface makes an equilibrium angle $\theta_e$ with the solid as a
result of intermolecular interaction between the three phases at the
contact line. Neglecting the contact angle hysteresis, $\theta_e$
takes a well-defined value determined by Young's law. The contact line
equilibrates at a distance $\Delta$ above the bath, which can be
expressed as
\begin{equation}\label{eq:theta}
\Delta = \pm \ell_\gamma \sqrt{2\left[ 1- \cos (\alpha - \theta_e)
    \right]}.
\end{equation}
Here, $\ell_{\gamma}=\sqrt{\frac{\gamma}{(\rho_{\ell}-\rho_{g}) g}}$
is the capillary length, defined by surface tension $\gamma$, gravity
$g$ and the density difference $\rho_\ell - \rho_g$. The $\pm$ sign
depends on whether $\theta_e$ is smaller (+) or larger (-) than the
plate inclination $\alpha$. When addressing the transition to air
entrainment, we will consider the upper phase A in
Fig.~\ref{fig1setup} to be gaseous, while phase B is a liquid. We
therefore use subscripts `$\ell$' and `$g$' to indicate liquid and gas
phase respectively.

When the plate is moving, the viscous drag generated by the moving
plate gradually deforms the fluid-fluid interface. As long as the
speed of the plate is lower than a threshold value, the meniscus
equilibrates at a new distance $\Delta$ from the bath level (see
Fig. \ref{fig1setup}). For positive $U_p$ the plate is moving
downwards so that $\Delta$ is lower than the static equilibrium
height, while the opposite holds for negative $U_p$. The contact line
is assumed to be straight so that the problem can be treated as
two-dimensional. The interface is described by the film thickness
profile $h(x)$. The origin $x=0$ is chosen at the contact line.  When
the plate is moving beyond a critical speed, the meniscus can no
longer equilibrate to a steady state. In the case of receding contact
lines (withdrawing plate) this leads to the deposition of a liquid
film~\cite{BR97,E04b,SP91,Q91,SDAF06, SADF07, DFSA07}, while air will
be entrained for advancing contact
lines~\cite{DYCB07,LRHP11,BR79,BK76b,SK00,Ben07,Ben10, MCSA12}.

In this study we only focus on viscous flows, for which the Reynolds
number is assumed to be zero. Thus for incompressible fluids, the flow
fields in the fluids are described by Stokes equations and continuity,
\begin{equation}\label{stoke1}
\begin{array}{l l l}
\eta_{g}\nabla^{2}\vec{u}_{g}-\vec{\nabla}
p_{g}-\vec{\nabla}\Phi_{g}=0,& \vec{\nabla}\cdot\vec{u}_{g}=0,
\end{array}
\end{equation}
and
\begin{equation}\label{stoke2}
\begin{array}{l l l}
\eta_{\ell}\nabla^{2}\vec{u}_{\ell}-\vec{\nabla}
p_{\ell}-\vec{\nabla}\Phi_{\ell}=0, &
\vec{\nabla}\cdot\vec{u}_{\ell}=0,
\end{array}
\end{equation}
where $\vec{u}_{g}$ and $\vec{u}_{\ell}$, $p_{g}$ and $p_{\ell}$,
$\Phi_{g}$ and $\Phi_{\ell}$ are the corresponding velocity fields,
pressures and gravitational potentials in phase A and phase B
respectively. When considering air entrainment, phase A is assumed to
be a gas of viscosity $\eta_g$, and phase B is a liquid of viscosity
$\eta_\ell$. The relative viscosity is expressed by the viscosity
ratio $R=\eta_g/\eta_\ell$ which, in practical situations, can be very
small.

To solve for the flow fields and the interface shape, we need to
specify appropriate boundary conditions. At the steady interface, the
velocities parallel to the interface $u^t$ are continuous and the
velocities normal to the interface $u^n$ vanish, so
\begin{equation}\label{flowBC1}
u_g^t=u_{\ell}^t
\end{equation}
and
\begin{equation}\label{flowBC2}
u_g^n=u_{\ell}^n=0.
\end{equation}
In each phase, we define the normal stress
$\sigma^n\equiv\hat{n}\cdot\widehat{\sigma}\cdot\hat{n}$ at the
interface, where $\hat{n}$ is the unit vector normal to the interface
and the stress tensor $\widehat{\sigma}$ is defined as (in cartesian
coordinates)
\begin{equation}\label{def_stress}
\sigma_{ij}\equiv-p\delta_{ij}+\eta\left (\frac{\partial u_i}{\partial
  x_j}+\frac{\partial u_j}{\partial x_i}\right).
\end{equation}
The normal stress discontinuity across the interface is related to the
curvature $\kappa$ and to the surface tension $\gamma$ by Laplace's
law
\begin{equation}\label{flowBC4}
\sigma_{\ell}^n-\sigma_{g}^n=\gamma \kappa.
\end{equation}
By contrast, the tangential stress component $\sigma^t
\equiv\hat{t}\cdot\widehat{\sigma}\cdot\hat{n}$ is continuous across
the interface,
\begin{equation}\label{flowBC3}
\sigma_g^t=\sigma_{\ell}^t.
\end{equation}
At the solid/fluid boundary ($y=0$), the velocity normal to the wall
$u_{y}(y=0)$ vanishes as no penetration of fluid through the solid is
allowed, i.e.
\begin{equation}\label{flowBC5}
u_{y}(y=0)=0.
\end{equation}
Regarding the velocity component parallel to the wall, $u_x(y=0)$, the
situation is more subtle because of the moving contact line
singularity~\cite{HS71,BEIMR09}: imposing a no-slip boundary condition
leads to diverging stress fields and calls for a microscopic mechanism
of regularization. In the following section we present two methods to
solve the flow equations, which naturally involve different
regularizations of the singularity: a generalized lubrication model
and Lattice Boltzmann simulations. The microscopic boundary condition
will therefore be discussed separately below.

\section{Methods}\label{methods}

In this section we present two methods to determine the meniscus shape
sketched in Fig.~\ref{fig1setup}. We first present a model that can be
considered as a generalization of the standard lubrication
approximation. We then present the Lattice Boltzmann method, which is
a rather different approach to solve for the flow and the meniscus
shape. The models will be refereed to as GL (Generalized Lubrication
model) and LB (Lattice Boltzmann).

\subsection{Generalized Lubrication model}\label{LTmodel}

\subsubsection{Derivation}\label{LTmodel_derivation}
\begin{figure}
\begin{center}
\includegraphics[width=0.4\textwidth]{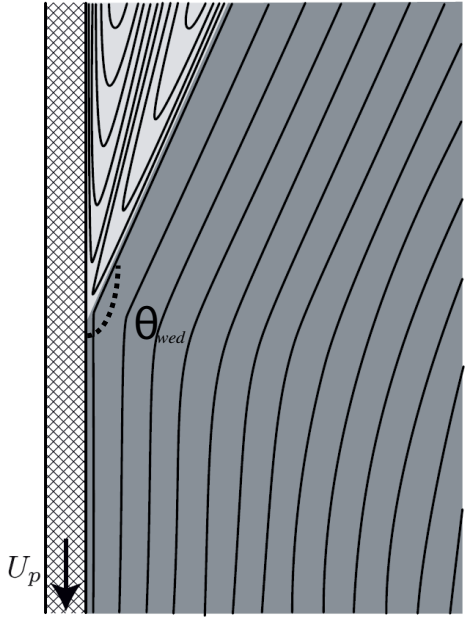}
\caption{\label{flow}Streamlines of the flow in a wedge of angle
  $\theta_{wed}$, which in this case is close to $\pi$. This flow
  solution has been derived analytically by Huh \&
  Scriven~\cite{HS71}, and is used here to derive a generalized
  lubrication model.}
\end{center}
\end{figure}

The lubrication approximation has been a very efficient framework to
deal with thin film flows \cite{ODB97}. This systematic reduction of
the Navier-Stokes equations is very suitable for numerical simulations
and in many cases allows for analytical results~\cite{BEIMR09}. It is
usually derived for flows that involve a single phase that constitutes
a ``thin'' film, i.e. the slopes $dh/dx$ are assumed small. However,
the expansion parameter underlying the analysis is not the interface
slope, but the capillary number ${\rm
  Ca}=U\eta_\ell/\gamma$~\cite{ODB97}. This means that a
lubrication-type analysis can be performed whenever surface tension
dominates over viscosity. Indeed, it was shown in~\cite{SN06} that the
lubrication approximation can be generalized to large interface angles
$\theta$, giving a perfect agreement with the perturbation results by
Voinov~\cite{V76} and Cox~\cite{C86}. Here we further extend this
approach by taking into account,  besides of the effect of a large slope, also
of the viscous flows on both sides of the
interface. The goal is to provide a model that can deal with moving
contact lines in cases where both phases are important (as in
Fig.~\ref{fig1setup}). In particular, this will allow us to study the
air entrainment transition.
Let us now derive this generalized lubrication model. As
mentioned in the preceding section, the interface curvature $\kappa$
is determined by the normal stress difference across the interface. In
curvilinear coordinates, we write
\begin{equation}\label{youglaplace2}
\gamma\kappa\equiv\gamma\frac{d\theta}{ds}=\sigma^n_{\ell}-
\sigma^n_{g},
\end{equation}
where $\theta$ is the local interface angle and $s$ the arc length
(see Fig. \ref{fig1setup}). The normal stresses have to be determined
from the flows in the fluids, which themselves depend on the shape of
the interface. For the usual lubrication theory in which the interface
slope is small, the leading order contribution to the flow reduces to
a parabolic Poiseuille flow inside the film. This can be generalized
to two-phase flows and large interface slopes: as long as the
capillary number is small, the interface curvature is small as well
and the leading order velocity field is given by the flow in a wedge
(Fig.~\ref{flow}). The wedge solutions have been obtained analytically
by Huh $\&$ Scriven~\cite{HS71}, for any viscosity ratio $R=\eta_g/\eta_{\ell}$
and for any wedge angle $\theta$. Fig. \ref{flow} shows the
corresponding streamlines. Based on these exact solutions, the local
normal stress can be determined, thus giving the local curvature of
the interface through Eq. (\ref{youglaplace2}).

We denote the quantities derived from the Huh-Scriven solutions by capital
symbols, e.g. normal stress is denoted by $\Sigma^n$, velocity by
$\vec{U}$ and pressure by $P$. For the Huh-Scriven solutions, it turns
out that the non-isotropic part of the normal stress at the interface
vanishes so that
\begin{equation}
\Sigma^n=-P.
\end{equation}
Approximating the normal stresses by the Huh-Scriven solutions,
Eq. (\ref{youglaplace2}) becomes
\begin{equation}\label{youglaplace3}
\gamma\frac{d\theta}{ds}=\Sigma^n_{\ell}- \Sigma^n_{g}=P_{g}-P_{\ell}.
\end{equation}
Since $P_{g}-P_{\ell}$ are defined up to an integration constant, it
is convenient to differentiate Eq. (\ref{youglaplace3}) once with
respect to $s$, giving

\begin{eqnarray}\label{int1}
\gamma\frac{d^{2}\theta}{ds^{2}}=\frac{dP_{g}}{ds}-\frac{dP_{\ell}}{ds}
=\left[\vec{\nabla} P_{g} - \vec{\nabla}P_{\ell} \right]_{\rm
  int}\cdot \hat{e}_{s}.
\end{eqnarray}
The index ''int'' indicates that the quantities inside the brackets
are to be evaluated on the interface and $\hat{e}_{s}=\hat{t}$ is the
unit vector tangent to the interface.

When Stokes equation (\ref{stoke1}) is applied, Eq. (\ref{int1}) can
be rephrased in terms of Huh-Scriven velocity fields
\begin{equation}\label{eq:hup}
\gamma\frac{d^{2}\theta}{ds^{2}}=\left[\eta_{g}\nabla^{2}\vec{U}_{g}-\eta_{\ell}\nabla^{2}\vec{U}_{\ell}-\vec{\nabla}(\Phi_{g}-\Phi_{\ell})\right]_{\rm
  int}\cdot \hat{e}_{s}.
\end{equation}
The viscous contributions on the right hand side can be expressed in
terms of $R$ and $\theta$ in the form %
\begin{equation}
\eta_{\ell}\left[\frac{\eta_{g}}{\eta_{\ell}}\nabla^{2}\vec{U}_{g}-\nabla^{2}\vec{U}_{\ell}\right]_{int}\cdot
\hat{e}_{s}=\frac{3\eta_{\ell} U_p f(\theta,R)}{h^{2}},
\end{equation}
where
\begin{eqnarray}\label{eq:f}
f(\theta,R)&\equiv& \frac{2\sin^{3}\theta[R^{2}f_1(\theta)+2R
    f_3(\theta)+f_1(\pi-\theta)]}{
  3[Rf_1(\theta)f_2(\pi-\theta)-f_1(\pi-\theta)f_2(\theta)]}
\nonumber\\ f_1(\theta)&\equiv&\theta^{2}-\sin^{2}
\theta\nonumber\\ f_2(\theta)&\equiv&\theta-\sin\theta
\cos\theta\nonumber\\ f_3(\theta)&\equiv&(\theta(\pi-
\theta)+\sin^{2}\theta).\nonumber\\
\end{eqnarray}
The gravity terms in (\ref{eq:hup}) can be simplified to
\begin{equation}
\vec{\nabla}\left[\Phi_{g}-\Phi_{\ell}\right]_{int}\cdot
\hat{e}_{s}=-(\rho_\ell-\rho_g)g\sin(\theta-\alpha),
\end{equation}
where $\rho_g $ and $\rho_{\ell}$ are the densities of the two phases.

The final result of the analysis is a generalized form of the
lubrication equation, which after scaling all lengths with the
capillary length
$l_{\gamma}=\sqrt{\frac{\gamma}{(\rho_{\ell}-\rho_{g}) g}}$ becomes
\begin{equation}\label{interface_eqn}
\frac{d^{2}\theta}{ds^{2}}=\frac{3\rm
  Ca}{h^{2}}f(\theta,R)+\sin(\theta-\alpha).
\end{equation}
In the reminder we will use the same symbols for rescaled
lengths. Once more, the capillary number is defined based on the
viscosity of the liquid, ${\rm Ca}=\eta_{\ell}U_p/\gamma$. This
equation must be complemented by the geometrical relation
\begin{equation}\label{geo}
\frac{dh}{ds}= \sin \theta.
\end{equation}
Note that for all numerical examples in the rest of the paper, we
consider a vertical plate, for which $\alpha=\pi/2$.

One easily verifies that the standard lubrication equation is
recovered when taking the limit of vanishing $R$, $\theta$ and
$\alpha$. Namely, $f(0,0)=-1$ and (\ref{interface_eqn}) reduces to
\begin{equation}\label{first_lubrication_eqn}
h'''=\frac{3\rm Ca}{h^2}-h'+\alpha.
\end{equation}
When considering a single phase with arbitrary angle, one recovers the
equation previously proposed in~\cite{SN06}, with $f(\theta,0)$
instead of $f(\theta,R)$.

\subsubsection{Slip boundary condition}\label{LTmodel_slip}
It is important to note that Eq.~(\ref{interface_eqn}) is derived from
the Huh-Scriven solution with no-slip boundary condition. Near the
contact line, however, this induces a divergence of the pressure and
of the shear stress, which scale as $\sim \eta_{\ell} U_p/h$ and $\sim
\eta_gU_p/h$ in the liquid and the gas respectively. A purely
hydrodynamic approach to regularize the singularity is to impose a
slip boundary condition on the solid wall. No analytical solution for
the flow in a wedge with slip can be obtained. However,
as contact line flows are only mildly affected by the details of the
microscopic conditions~\cite{E04a,BEIMR09}, we proceed by a
phenomenological regularization. We therefore consider the standard
lubrication equation, which can be derived including a Navier slip
boundary condition. It reads
\begin{equation}\label{lubrication_eqn}
h'''=\frac{3\rm Ca}{h(h+3\lambda_s)}-h'+\alpha,
\end{equation}
where $\lambda_s$ is the slip length. In comparison to
(\ref{first_lubrication_eqn}), the effect of slip can be summarized by
a correction factor $h/(h+3\lambda_s)$ for the viscous term. Indeed,
this weakens the singularity such that the equations can be integrated
to $h=0$ \cite{BSB03}.
We simply propose to use the same regularization factor for the
viscous term in (\ref{interface_eqn}),  i.e.
\begin{equation}\label{full_interface_eqn}
\frac{d^{2}\theta}{ds^{2}}=\frac{3\rm
  Ca}{h(h+3\lambda_s)}f(\theta,R)+\sin(\theta-\alpha),
\end{equation}
where we have assumed the slip length to be independent of $R$.  The
appropriate boundary conditions are that the equilibrium contact angle
$\theta_e$ is recovered at the contact line, and that the interface
attains the angle of the reservoir at infinity:
\begin{equation}\label{BC_interface}
\begin{array}{l l l l}
h(s=0)=0; & \theta(s=0)=\theta_{e}; &
\theta(s\rightarrow\infty)=\alpha.
\end{array}
\end{equation}
The meniscus shape is now fully determined by the lubrication equation
(\ref{full_interface_eqn}), geometry (\ref{geo}) and boundary
conditions (\ref{BC_interface}). For a given value of the capillary
number ${\rm Ca}$, the model parameters are the viscosity ratio $R$,
the contact angle $\theta_e$, and the microscopic length
$\lambda_s$. Below, we compute the shape of the meniscus for different
parameters by numerical integration, using a 4th order Runge-Kutta
numerical scheme. As the boundary conditions are imposed at different
locations, the solution is determined using a shooting algorithm.

\subsection{Lattice Boltzmann Method}
In this section we discuss some of the key features of the LB method. In a LB
 model, the following discrete
Boltzmann equation is solved for the single particle distribution
function $ f_{i,\alpha}({\bf x},t)$ over a 2D square lattice:
\begin{eqnarray}\label{eq:lbe} 
f_{i,\alpha} ({\bf x} + {\bf e}_i\Delta t,t +\Delta t) -
f_{i,\alpha}({\bm x},t) &=& \Omega_{i,\alpha}({\bf x},t)~,
\end{eqnarray}
where
$$\Omega_{i,\alpha}({\bf x},t) = -
\frac{1}{\sigma_\alpha}[f_{i,\alpha}({\bf x},t ) -
  f^\mathrm{eq}_{i,\alpha}({\bf x},t )],$$ is the single time
relaxation, linear BGK collision operator \cite{Bhatnagar1954}, $
f^\mathrm{eq}_{i,\alpha}({\bf x},t) $ is the discrete Maxwell
distribution function defined as:
\begin{equation}\label{eq:feq}
f^\mathrm{eq}_{i,\alpha}({\bf x},t) = \rho_\alpha w_i
\left[1+\frac{{\bf e}_i \cdot {\bf u}_\alpha}{c_s^2} + \frac{{\bf
      e}_i\cdot {\bf u}_\alpha}{2c_s^4} - \frac{{\bf u}_\alpha\cdot
    {\bf u}_\alpha}{2c_s^2}\right],
\end{equation}
where
\begin{equation}
\rho_\alpha =\sum_{i} f_{i,\alpha}, \hskip 0.3cm \rho_\alpha{\bf
  u}_\alpha= \sum_{i} f_{i,\alpha}{\bf e}_i~,
\label{eq:hydro}
\end{equation}
and $w_i$, ${\bf e}_{i}$ in (\ref{eq:feq}) are the weights and the
corresponding lattice speeds respectively
\cite{succi_book2001,Wolf-gladrow2005}.  The weights, $w_i$,
corresponding to the 2-dimensional and 9 velocity model, D2Q9, are
given by $w_0=4/9$, $w_1=w_2=w_3=w_4=1/9$, and
$w_5=w_6=w_7=w_8=1/36$. The total fluid density is $\rho=\sum_\alpha
\rho_\alpha$ and the total hydrodynamic velocity is ${\bf
  u}=\sum_{\alpha} \rho_\alpha{\bf u}_{\alpha}/\rho$.  The effective
kinematic viscosity is related to the relaxation time of the different
components $\nu=\sum_{\alpha} c_s^2 (\sigma_\alpha c_\alpha-0.5)$
\cite{Shan1995}, $c_\alpha=\rho_\alpha/\rho$ is the concentration, and
$c_s=1/\sqrt{3}$ is the speed of sound on the lattice. In absence of
an external force, each component satisfies the ideal gas equation of
state $p=c_s^2 \rho$. For multicomponent simulation we are using two
distribution functions ($\alpha = 1,2$), whereas for multiphase
simulations we restrict ourselves to only one distribution function
($\alpha = 1$).

\subsubsection{Multiphase/multicomponent model}
The multicomponent/multiphase algorithm is based on a standard
Shan-Chen lattice Boltzmann method \cite{Shan1993,Shan1994,Shan1995}.
The non-ideal nature of the fluid is introduced by adding an internal
force and shifting the lattice Boltzmann equilibrium velocity as
\begin{equation} 
{\bf u}^\mathrm{eq}_\alpha = {\bf u}^{\prime} +
\frac{\sigma_{\alpha}{\bf F}^\alpha}{\rho_\alpha},
\;\;\;\mbox{where}\;\;\; {\bf u}^\prime=\frac{\displaystyle
  \sum_{\alpha} \rho_\alpha {\bf
    u}_\alpha/\sigma_\alpha}{\displaystyle \sum_\alpha
  \rho_\alpha/\sigma_\alpha}.
\end{equation} 
For the non-ideal interaction the force ${\bf F}_{\alpha}$ in the
Shan-Chen model ~\cite{Shan1994, Shan1995} is given by:
\begin{equation}\label{eq:scforce}
{\bf F}_{\alpha} = -G_{\alpha\alpha^\prime} \psi_\alpha ({\bf x})
\sum_{i,\alpha\neq\alpha^\prime} w_i \psi_{\alpha^\prime} ({\bf x} +
    {\bf e}_{i}) {\bf e}_{i}~
\end{equation}
where $\{\alpha,\alpha^{\prime}\}=\{1,2\}$ are indices for two fluid
components while the coupling parameter $G_{\alpha\alpha^\prime}$ is
the strength of the interaction and determines the surface tension in
the model. This force allows for the spontaneous formation of an
interface between the different fluid components, i.e. no interface
tracking is needed. For multicomponent simulations $G_{12} = G_{21} =
G,~G_{11} = G_{22} = 0$, $\psi_\alpha = \rho_\alpha$. In the case of
multiphase simulations $\alpha = 1$, $\psi = 1- \exp(-\rho)$.  The
equation of state is modified to $p=c_s^2(\rho_1+\rho_2) + G c_s^2
\rho_1 \rho_2$ and $p=c_s^2\rho + \frac{G}{2} c_s^2 \psi^2$
respectively for multicomponent and multiphase simulations, where the
first term correspond to the ideal gas and the second term is the
non-ideal part due to the external Shan-Chen force.  Many validation
studies exist, showing that the hydrodynamical fields, ${\bf u}({\bf
  x},t), \rho({\bf x},t)$ satisfies the Navier-Stokes equations with a
non-ideal Pressure tensor, under a suitable multiscale Chapman-Enskog
expansion.

\subsubsection{Boundary conditions for LB simulations}
The no-slip boundary condition for the fluid corresponds to the bounce
back boundary condition \cite{succi_book2001} for the distribution
functions $f_{i,\alpha}({\bf x},t)$ defined at the boundary nodes.

The surface wetting for multicomponent simulations is introduced by
adding an additional force at the wall~\cite{Benzi2006,Huang2007}
\begin{equation} {\bf F}^{ads}_{\alpha} = -G_\alpha^{ads}
  \rho_{\alpha} ({\bf x},t) \sum_{i} w_i s({\bf x} + {\bf e}_{i}) {\bf
    e}_{i},
\end{equation}
where, $s({\bf x} + {\bf e}_{i})=1$ for a wall node and is $0$ for a
fluid node. The parameter $G_\alpha^{ads}$ can be varied to control
the wetting properties of the wall; in all our simulations we have
used $G_1^{ads} = -G_2^{ads}$. 
Similarly for multiphase simulations we fix a wetting parameter,
$\rho_{w}$ for the nodes in the wall and calculate the Shan-Chen force
(\ref{eq:scforce}) at the wall~\cite{Benzi2006}
\begin{equation} {\bf F}^{ads} = -G
  \psi (\rho_{w}) \sum_{i} w_i s({\bf x} + {\bf e}_{i}) {\bf e}_{i},
\end{equation}
where, $s({\bf x} + {\bf e}_i)=1$ for a wall node and is $0$ for a
fluid node. The parameter $\rho_{w}$ is varied to control the
equilibrium contact angle at the wall.
Let us stress that all LB methods, independent of the underlying
kinetic model, describe the multi-phase or the multi-component
dynamics via a {\it diffuse interface} approach. There exist therefore
a natural regularizing microscopic length which is of the order of the
interface width, typically a few grid points.  Such a length scale is
also of the order of the {\it effective} slip length 
that must be used whenever a quantitative comparison between the
hydrodynamical behavior of the LB and the evolution of the equivalent
Navier-Stokes system is made, as shown for example recently in
\cite{Sbragaglia2008}.

\section{Comparing the lubrication  model and Lattice Boltzmann simulations}\label{compare}
In this section we compare the results of the GL model and the LB
simulations. Since the latter are limited to moderate viscosity ratios
$R$, the comparison is done for $R = 0.03, 0.8$ and 1. We further
explore the parameter space in Sec.~\ref{sec.GL}, using only the
lubrication approach.

The results for this section are computed for $\theta_e=\pi/2$ and
$\alpha=\pi/2$. The lattice separation in LB is 0.01 (in capillary
length units), which will be related to an effective slip length from the comparison with the lubrication model.

\subsection{Meniscus rise}
\begin{figure}
\begin{center}
\includegraphics[width=0.7\textwidth]{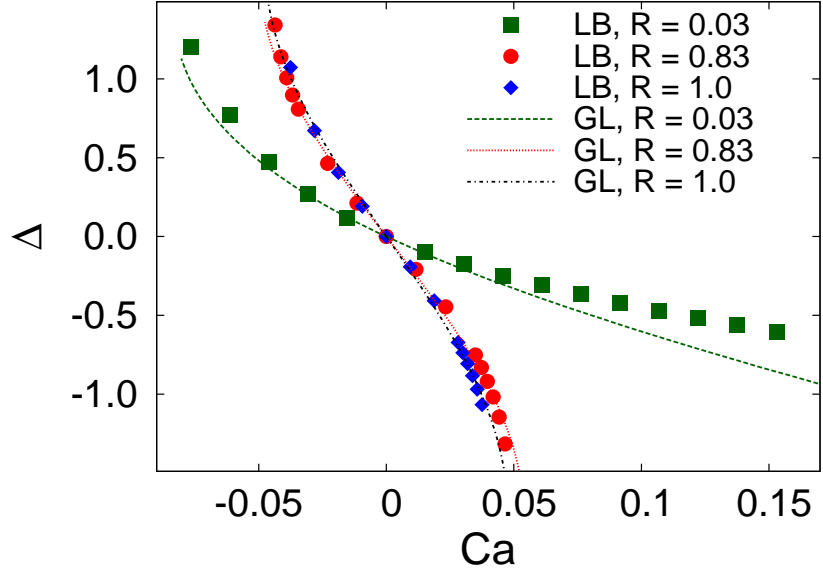}
\caption{\label{delta_ca_compare}(Color online) Meniscus deformation
  $\Delta$ as a function of $\rm Ca$ for $\theta_e=\pi/2$. $\rm Ca$ is
  negative/positive when the plate is moving upward/downward. Symbols:
  results of lattice Boltzmann (LB) simulation. Lattice separation
  $=0.01$. Curves: results of the lubrication-type (GL) model with a
  slip length $\lambda_s = 0.002$. All lengths are scaled by the
  capillary length $\ell_\gamma$. }
\end{center}
\end{figure}

We first compare the meniscus rise $\Delta$ for viscosity ratios $R=0.03, 0.8$ and 1 in
Fig. \ref{delta_ca_compare}. When the plate is at rest, ${\rm Ca}=0$,
the meniscus is perfectly horizontal $\Delta=0$ due to the choice of
$\theta_e$ and $\alpha$. Let us first consider the case where both
liquids have identical viscosity, $R=1$, but are still immiscible due
to the nonzero surface tension. This case is perfectly symmetric in
the sense that $\rm Ca \rightarrow -\rm Ca$ gives $\Delta \rightarrow
-\Delta$: there is no difference between plunging and
withdrawing. This symmetry is indeed observed in
Fig. \ref{delta_ca_compare}. Blue diamonds represent LB simulations,
while the dash-dotted line is the GL model. We use this symmetric case
to calibrate the microscopic parameter of the GL model. A nearly
perfect fit is achieved for slip length $\lambda_s=0.002$, which is a
reasonable value given that the grid size used in the LB simulation is
0.01. As $\rm Ca$ increases, the viscous forces increasingly deform
the interface, leading to a change in $\Delta$.

It is interesting to see to what extent the same microscopic parameter
$\lambda_s$ is able to describe different viscosity ratios. We first
mildly decrease the viscosity ratio to $R=0.83$ and still find a very
good agreement between LB an GL (red circles and dotted line
respectively). With respect to the case $R=1$ we see that $|\Delta|$ is
slightly smaller at a given value of $\rm Ca$. This means that for the
same speed, the meniscus is deformed by a smaller amount due to the
lower viscosity of the upper phase. When further decreasing the
viscosity ratio to $R=0.03$ (green squares, dashed line), some
differences between LB and GL becomes apparent (the same value for
$\lambda_s$ is maintained). The meniscus in GL is systematically below
the value obtained in LB. A possible explanation for this difference
is the sensitivity of the result on the microscopic contact angle
imposed as a boundary condition. Still, both models agree reasonably
well and display very similar trends. In particular, we find that much
larger values of $\rm Ca$ can be achieved due to the strong reduction
of the viscosity in the upper phase. This effect is most pronounced
for the plunging case, for which the liquid is advancing. This is
consistent with experimental observations that advancing contact lines
can move much more rapidly than receding contact lines
\cite{BEIMR09,Snoeijer13,MCSA12}. The breakdown of the steady solutions, which signals the
transition to air/liquid entrainment, will be discussed in details in
Sec.~\ref{sec.GL}.

\subsection{Shape of the meniscus}
\begin{figure}
\begin{center}
\includegraphics[width=0.55\textwidth]{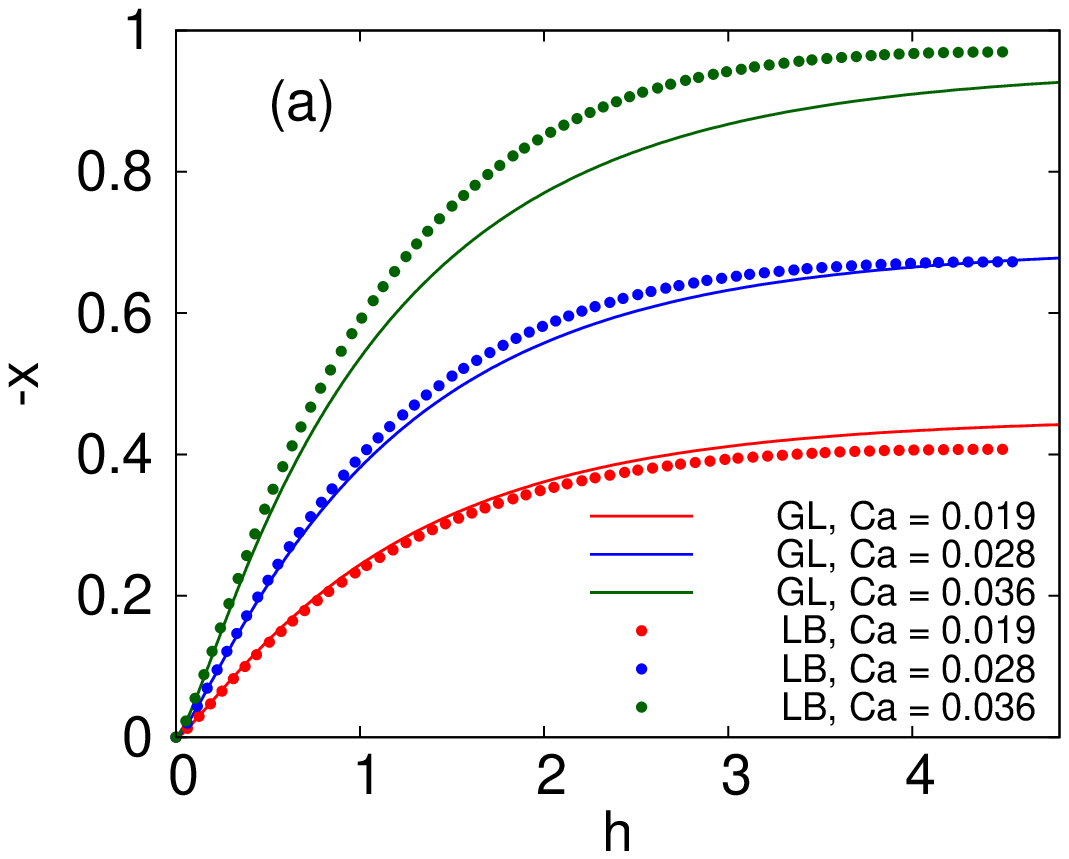}
\includegraphics[width=0.55\textwidth]{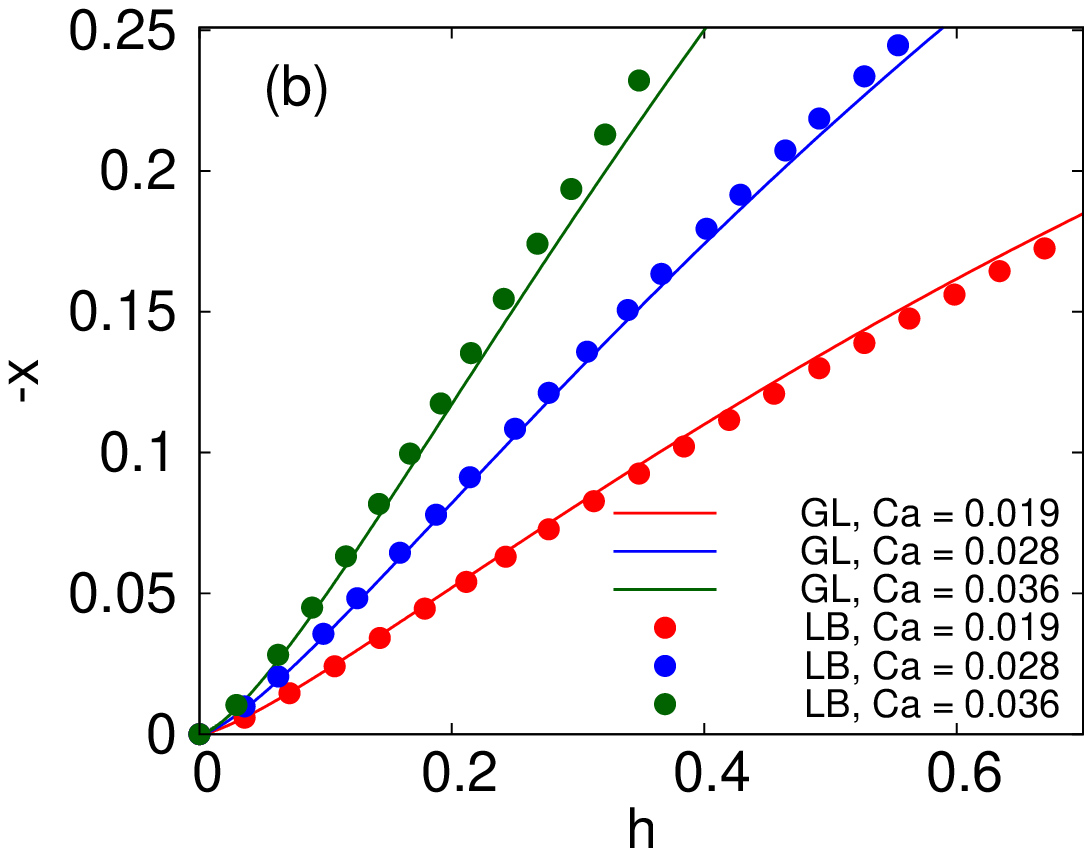}
\caption{\label{compare_profile}(Color online) Dynamical meniscus
  profiles $-x$ vs $h$ for $R=1$ ($\theta_e=\pi/2$ and
  $\lambda_s=0.002$ for the GL model, lattice separation $= 0.01$ for
  the LB simulation). All lengths have been scaled by the capillary
  length. The contact line is at $x=0$ so that the bath is at
  different $x$ for different $\rm Ca$. (a) Solid curves: results of
  GL model. Dots: results of LB simulation. (b) Zoom
  on the contact line region. }
\end{center}
\end{figure}

A much more detailed test for the two models is to investigate the
detailed structure of the interface: How well do the shapes of the
menisci compare between GL and LB? In Fig. \ref{compare_profile}, we
show the dynamical meniscus profiles for $\rm Ca$ = 0.019, 0.028 and
0.036, in the case of equal viscosities, $R=1$. Note that the contact
line position is held constant at $x=0$, so that the bath appears at
different heights due to the increase in magnitude of $\Delta$ with
speed. The agreement of the results of GL model and LB simulation is
very good in particular for $\rm Ca$=0.019 and 0.028, even down to the
contact line region (Fig. \ref{compare_profile}b). For larger plate
velocities some differences become apparent. These differences may be
due to different reasons. First, one has to notice that a large $\rm Ca$
is also accompained by a large viscous stress contribution, leading to
a larger bending of the interface and therefore to the possibility to
leave the realm of application of the GL model. Second, as said, in
the LB approach the {\it effective} slip length is an output and not
an input as for the GL, and it is not clear that one would need a
finer tuning of it as a function of the capillary number, in order to
match the GL behavior.

\begin{figure}
\begin{center}
\includegraphics[width=0.6\textwidth]{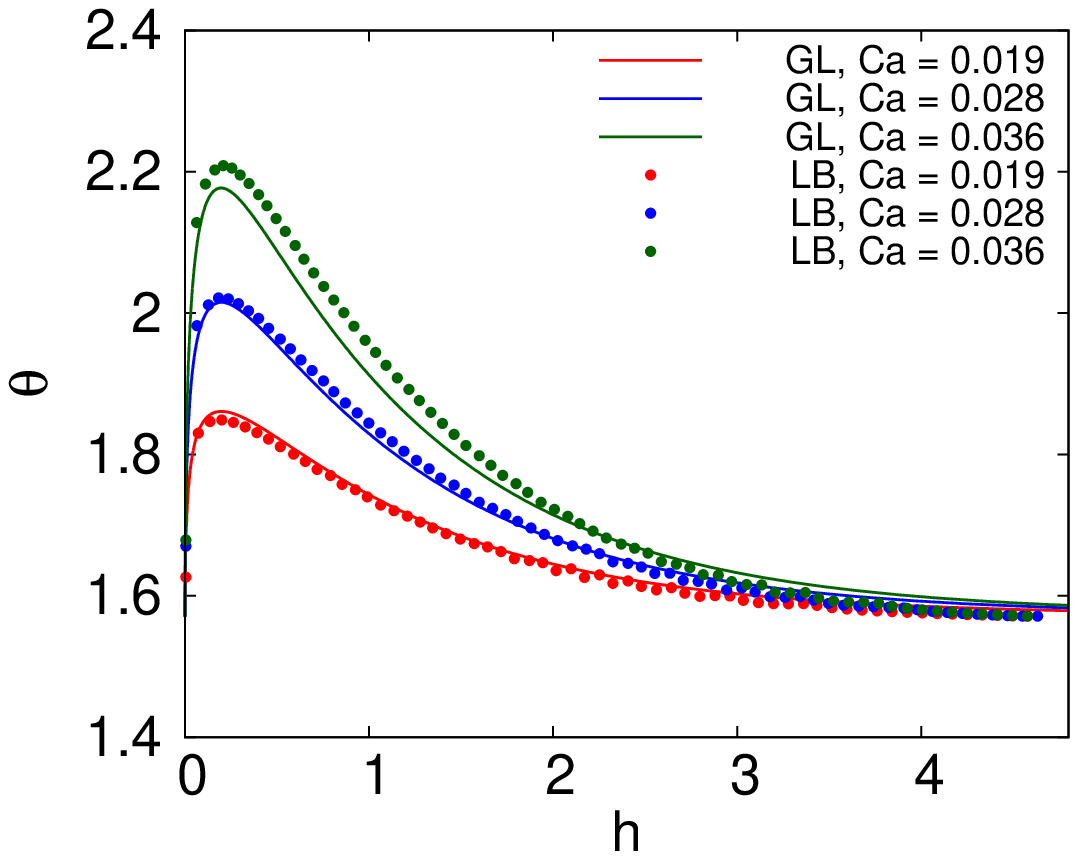}
\includegraphics[width=0.6\textwidth]{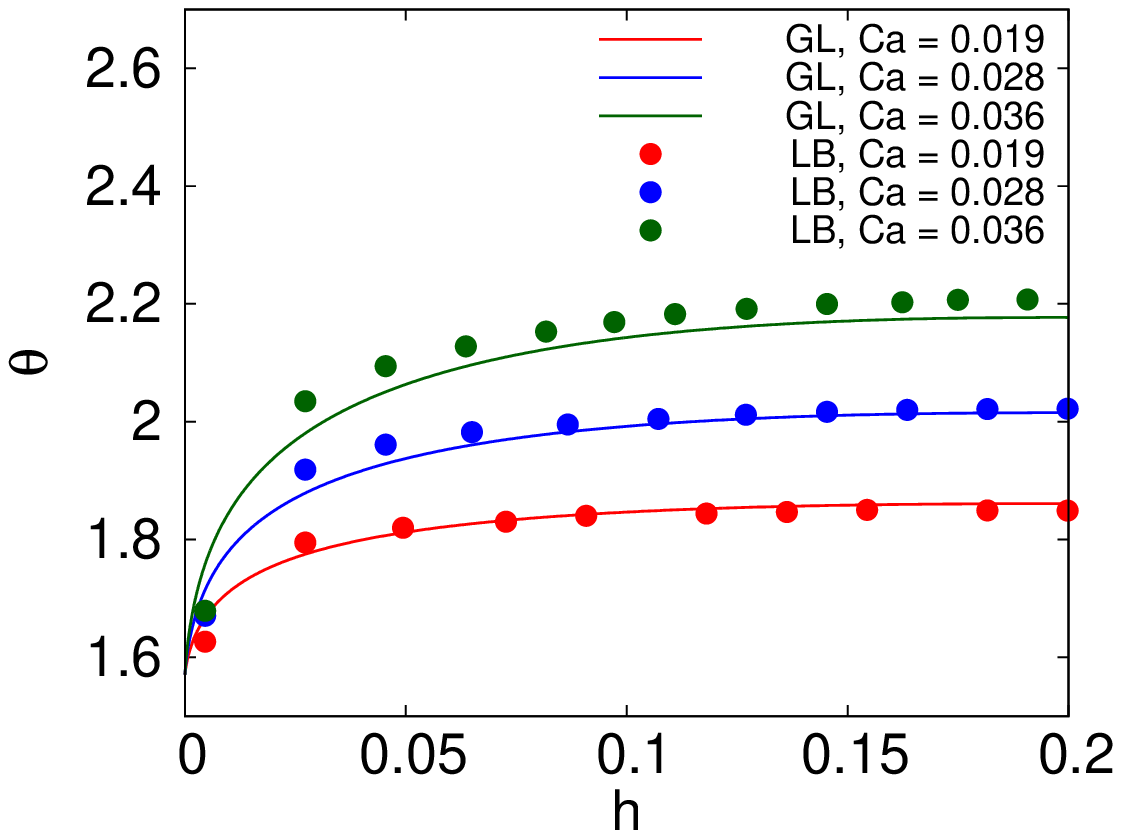}
\caption{\label{theta_h}(Color online) Local meniscus angle $\theta$
  vs $h$ for $R = 1$. Identical parameters as in
  Fig.~\ref{compare_profile}.}
\end{center}
\end{figure}

An even more detailed characterization of the meniscus shape is
provided by the local angle of the interface $\theta$ vs $h$, see
Fig. \ref{theta_h}. Clearly, both the GL model and the LB simulation
exhibit the same nontrivial variation of the contact angle. At small
scales, the angle approaches $\theta_e=\pi/2$, while at large scale
the meniscus evolves towards the reservoir $\theta=\alpha=\pi/2$. In
between, the angle changes significantly due to the well-known effect
of ``viscous bending'' \cite{BEIMR09}: the balance of viscosity and
surface tension leads to a curvature of the interface. Very similar
variations of the meniscus angle have been obtained experimentally
\cite{RG96}.

\section{Maximum speeds and transition to air entrainment}\label{sec.GL}
In this section we discuss the physics of air entrainment in the case
of a plunging plate ($\rm Ca>0$). For realistic situations the
viscosity ratio $R$ is typically very small, of order $10^{-2}$ for
water and much smaller for very viscous liquids. This regime can be
accessed by the GL model only, as LB is restricted to moderate
viscosity contrasts. In the first part of this section we discuss how
the transition to air entrainment is captured in our model in terms of a bifurcation
diagram. Next we study the effect of viscosity ratio on the critical speed. In the last part we investigate how the critical speed
depends on the microscopic parameters such as the slip length and the
static contact angle.

\subsection{Maximum speed for advancing contact lines}
We first consider a case where the equilibrium contact angle is close
to $\pi$. For such a hydrophobic substrate we expect air entrainment
to occur at relatively small ${\rm Ca}$ \cite{DYCB07,LRHP11}, and
therefore relatively weak curvature of the interface. This is
important, since the assumption underlying the GL model is that the
interface curvature is weak \cite{SN06}. We will therefore focus on
$\theta_e=2.8$ radians and explicitly verify how strongly the
interface is curved for our numerical solutions.
\begin{figure}
\begin{center}
\includegraphics[width=0.65\textwidth]{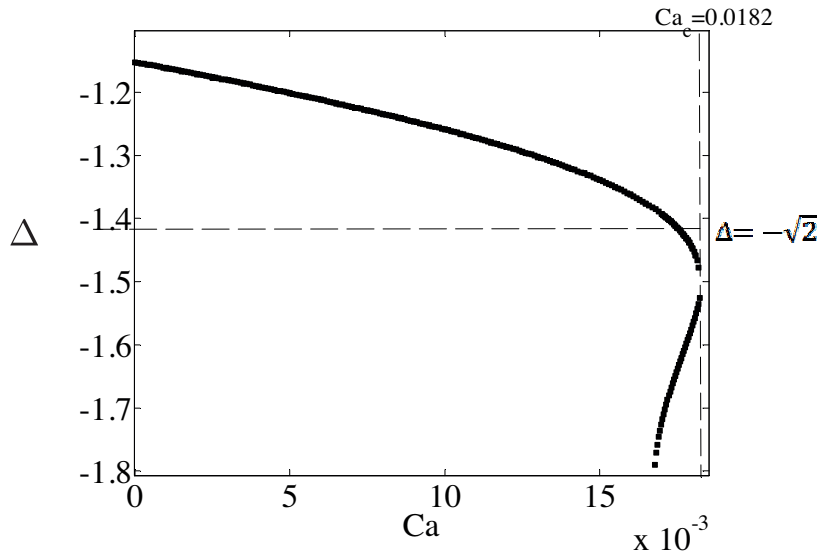}
\caption{\label{R0_01_DelCa}Meniscus fall $\Delta$
  versus $\rm Ca$ ($\theta_e=2.8$ radians, $R=0.01$,
  $\lambda_s=10^{-5}$). The horizontal dashed line indicates the
  minimum value of $\Delta$ for a static meniscus ($=-\sqrt{2}$, with
  $\theta_e=\pi$).}
\end{center}
\end{figure}
Figure~\ref{R0_01_DelCa} shows the drop of the meniscus $\Delta$ as
function of $\rm Ca$ for a viscosity ratio $R=0.01$
($\lambda_s=10^{-5}$, i.e. of the order of 10~nm). As $\rm Ca$
increases, the contact line equilibrates at a lower position resulting
in a more negative value of $\Delta$. However, when $\rm Ca$ achieves
a certain critical value, stationary solutions cease to exist. This
corresponds to a maximum plate velocity, or critical capillary number
$\rm Ca_c$. By analogy to deposition of liquid films for plate
withdrawal \cite{SADF07,DFSA07, ZSE09,SDAF06, SZAFE08}, this can be
identified as the onset of air entrainment: above ${\rm Ca}_c$,
unsteady solutions will develop, with a downward motion of the contact
line ~\cite{MCSA12}. As can be seen in Fig.~\ref{R0_01_DelCa}, the
critical point arises close to $\Delta=-\sqrt{2}$, which according to
(\ref{eq:theta}) corresponds to a meniscus with apparent contact angle
$\pi$. This is the analogue of the withdrawal case, for which $\Delta
= + \sqrt{2}$ and the apparent contact angle vanishes at the
transition \cite{E04b,SADF07}. Note that viscous effects push system
slightly below this maximum extent of deformation for a perfectly
static meniscus, with the critical point slightly below $\Delta =
-\sqrt{2}$.

\begin{figure}
\begin{center}
\includegraphics[width=0.65\textwidth]{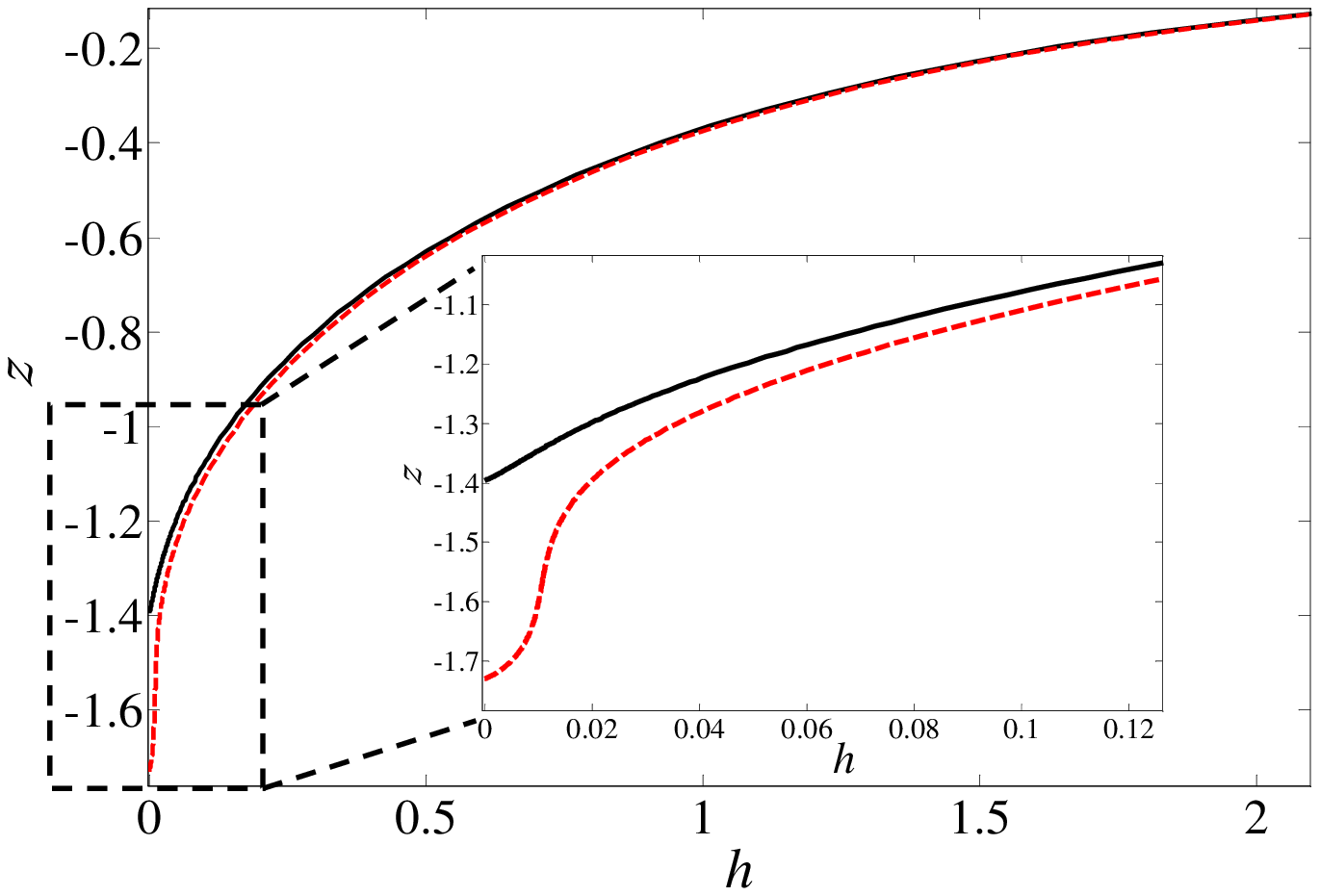}
\caption{\label{twoprofile}(Color online) Meniscus profiles for the
  upper branch (black solid curve) and the lower branch (red dashed
  curve) solutions for $\rm Ca=0.017$ ($\theta_e=2.8$ radians,
  $R=0.01$, $\lambda_s=10^{-5}$). Here we define $z=\Delta-x$. So the
  bath level is at $z = 0$.}
\end{center}
\end{figure}

Interestingly, for a range of speeds $\rm Ca<\rm Ca_c$ one actually
finds more than one solution (cf. Fig. \ref{R0_01_DelCa}). Upon
decreasing $\Delta$, the capillary number first increases and then
decreases close to the critical point. We refer to the solution
branches around ${\rm Ca}_c$ as the upper and lower branch
respectively. Once again, an identical bifurcation structure was
previously observed for the withdrawing plate
case~\cite{SADF07,CGS11,CSE12}. To compare these two types of solutions, we
show the corresponding meniscus profiles for $\rm Ca=0.017$ in
Fig. \ref{twoprofile}. At a large distance from the contact line, the
solutions are almost identical in shape. Zooming in on the contact
line region, however, we see the lower branch (red dashed curve)
solution displays a ``finger'' that explains the larger magnitude of
$\Delta$ with respect to the upper branch.

\begin{figure}
\begin{center}
\includegraphics[width=0.65\textwidth]{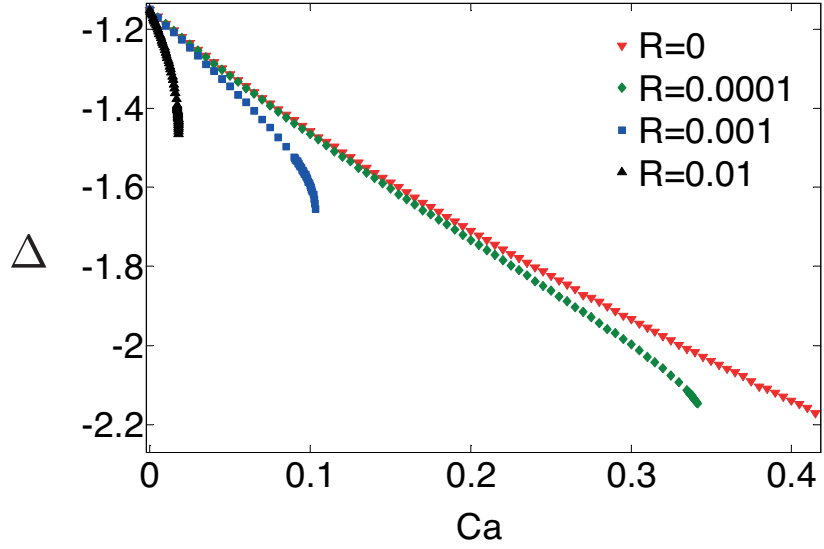}
\caption{\label{deltaca}(Color online) Meniscus fall $\Delta$ versus $\rm Ca$ for
  different viscosity ratios ($\theta_e=2.8$ radians,
  $\lambda_s=10^{-5}$). For the case the gas phase has no viscosity,
  $R=0$, steady-state menisci can be maintained to arbitrarily large
  velocity (within our numerical resolution).}
\end{center}
\end{figure}

\begin{figure}
\begin{center}
\includegraphics[width=0.65\textwidth]{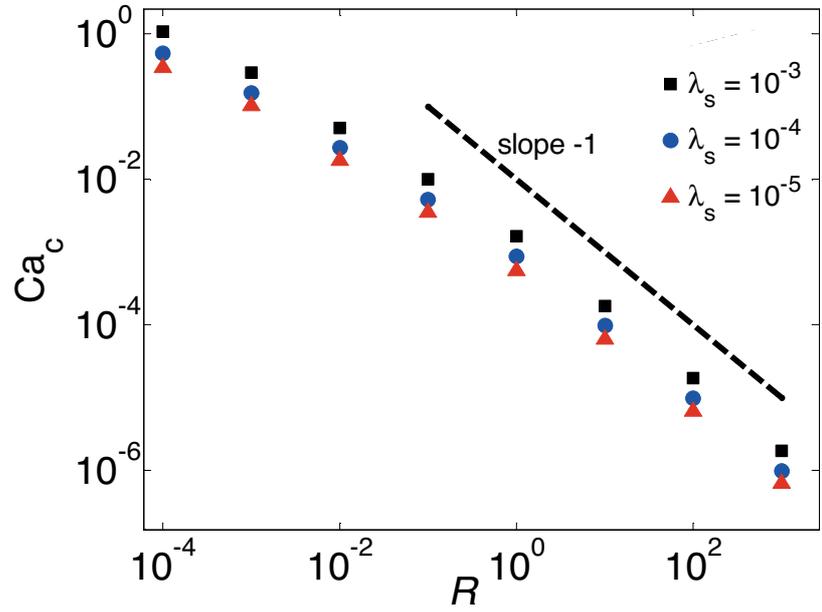}
\caption{\label{Ca_R}(Color online) Critical capillary number $\rm
  Ca_c$ versus viscosity ratio $R$ for different slip lengths
  $\lambda_s$ ($\theta_e=2.8$ radians). The dashed line indicates a
  power law with exponent -1, which is valid for large $R$.}
\end{center}
\end{figure}

\subsection{Effect of viscosity ratio}
A key parameter for the transition to air entrainment is the viscosity
ratio $R$. Figure~\ref{deltaca} shows the meniscus fall $\Delta$ as
function of $\rm Ca$ for different viscosity ratios: $R=0$, $10^{-4}$,
$10^{-3}$ and $10^{-2}$.  At $\rm Ca=0$, all cases have the same value
of $\Delta=-1.15$ corresponding to a static bath with contact angle
$\theta_e=2.8$. Now we consider the result for $R=0$, for which there
are no viscous effects in the air ($\eta_g=0$). We observe that
$\Delta$ decreases with $\rm Ca$, but without any bifurcation. It
appears that steady meniscus solutions can be sustained up to
arbitrarily large plate velocities. In fact, the curve is consistent
with the scaling $\Delta \sim -\sqrt{\rm Ca}$ at relatively large ${\rm Ca}$, corresponding to a
simple balance between gravity and viscosity. For $R\neq 0$, however,
the situation becomes fundamentally different. While the curves follow
the same trend as for $R=0$ at small ${\rm Ca}$, a deviation appears
at larger speeds that ultimately leads to a critical point. Each
nonzero viscosity ratio has a well-defined critical speed, with ${\rm
  Ca}_c$ increasing when the viscosity ratio $R$ is reduced.

These observations can be interpreted as follows. As long as the
viscosity of the air has a negligible effect on the flow, the curves
are indistinguishable from the case $R=0$. Deviations of the $R=0$
curve signal that the air flow starts to influence the shape of the
meniscus. Physically, this arises because the interface slope
approaches $\pi$, leaving only a narrow wedge angle for the air
flow. Figure~\ref{flow} illustrates that the recirculation in the air
induces significant velocity gradients: despite the small air
viscosity, the stresses in the small wedge of air become comparable to
those in the liquid. Mathematically, this can be derived from the
function $f(\theta,R)$ as defined in (\ref{eq:f}). For small $R$ and
$\theta$ close to $\pi$, we can approximate:
\begin{equation}\label{f_function}
f(\theta,R) \simeq f(\theta,0) -4R \simeq -\frac{2 (\pi -
  \theta)^3}{3\pi} -4 R
\end{equation}
as long as $\pi-\theta\gg2\pi R$. For $\theta$ very close to $\pi$,
$f(\theta,R)$ has a different asymptotic form, see the appendix for
details. The first term in (\ref{f_function}) represents the
(relative) viscous contribution inside the liquid, which vanishes in
the limit $\theta \rightarrow \pi$. The second term represents the
viscous contribution in the air, which will be significant once
$(\pi-\theta) \sim R^{1/3}$. Noting that the contact line is
\emph{receding} from the point of view of the air phase, one
understands that a critical speed appears when the effect of the air
becomes important.

The dependence of the critical speed $\rm Ca_c$ on the viscosity ratio
$R$ is shown in Fig.~\ref{Ca_R} (for various slip lengths). First we
consider the limit $R\gg1$, for which the upper fluid is actually much
more viscous than the bottom fluid. This is the usual case of a
receding contact line that is completely dominated by the upper
(receding) phase. In this limit we expect the critical speed to scale
with the viscosity of the upper phase, denoted $\eta_g$, such that
$U_c \sim \gamma/\eta_g$. Since we have based the capillary number on
the viscosity $\eta_\ell$, we obtain ${\rm Ca}_c \equiv
U_c\eta_\ell/\gamma \sim R^{-1}$. This is indeed observed in
Fig.~\ref{Ca_R} at $R\gg 1$. However, our main interest here lies in
the opposite limit, i.e. $R\ll1$, as for air entrainment. As already
mentioned, the critical speed seems to increase indefinitely by
reducing the viscosity ratio. This suggests that for the limiting case
of $R=0$, steady menisci can be sustained at arbitrarily large
plunging velocities. Our numerical resolution does not allow for a
perfect determination of the asymptotics for $R\ll 1$. Enforcing a
power law fit, ${\rm Ca}_c \sim R^{\beta}$, in the range
$R=10^{-4}-10^{-1}$, one obtains $\beta=-0.67$. This (effective)
exponent suggests that both phases play an important role in
determining the critical speed. Namely, the exponent would be
$\beta=-1$ if only $\eta_g$ were important, while $\beta=0$
corresponds to the case where $\eta_\ell$ is the only relevant
viscosity.

\begin{figure}
\begin{center}
\includegraphics[width=0.65\textwidth]{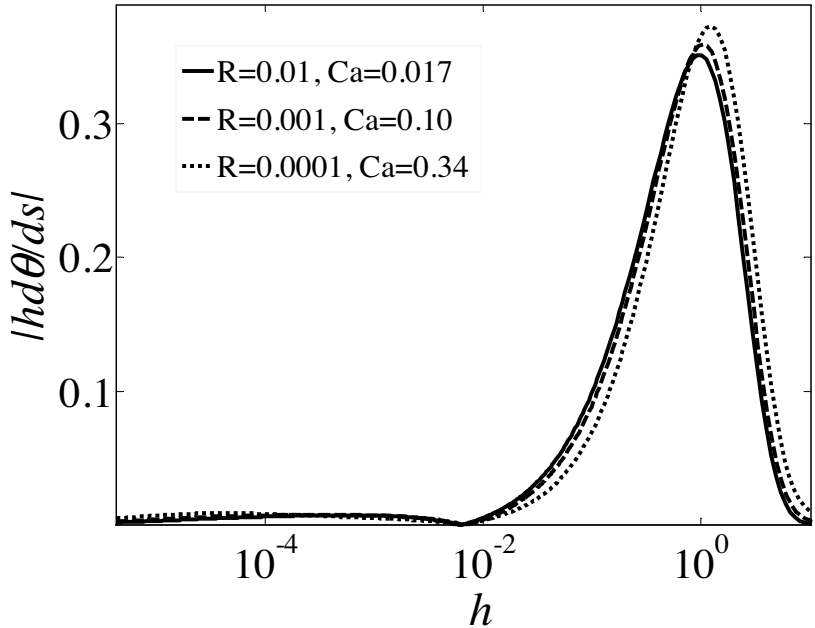}
\caption{\label{curvature}Scaled curvature $|hd\theta/ds|$ vs $h$ for
  $\rm Ca$ very close to $\rm Ca_c$ ($\lambda_s=10^{-5}$).}
\end{center}
\end{figure}

Finally, we briefly verify the assumption of small curvature,
necessary for the strict validity of the model. In
Fig. \ref{curvature} we plot the dimensionless curvature, $h
|d\theta/ds|$, as function of $h$ in the vicinity of the critical
point ($R=10^{-5},10^{-4}$ and $10^{-3}$, $\lambda_s=10^{-5}$). At
small scales, $h|d\theta/ds| \ll 1$ for all $\rm Ca$, consistent with
the assumption of small curvature. However, the curvature increases
significantly when approaching the bath due to the bending of
interface from a large contact angle to $\pi/2$. The magnitude is
acceptable in this regime, in particular since viscous effects becomes
less important at large scales. Inclining the plate angle to values
close to $\pi$ would further reduce this bending effect, and extend
the range of validity of the GL model.

\subsection{Dependence of the critical speed on microscopic parameters}\label{sec:parameters}

\begin{figure}
\begin{center}
\includegraphics[width=0.65\textwidth]{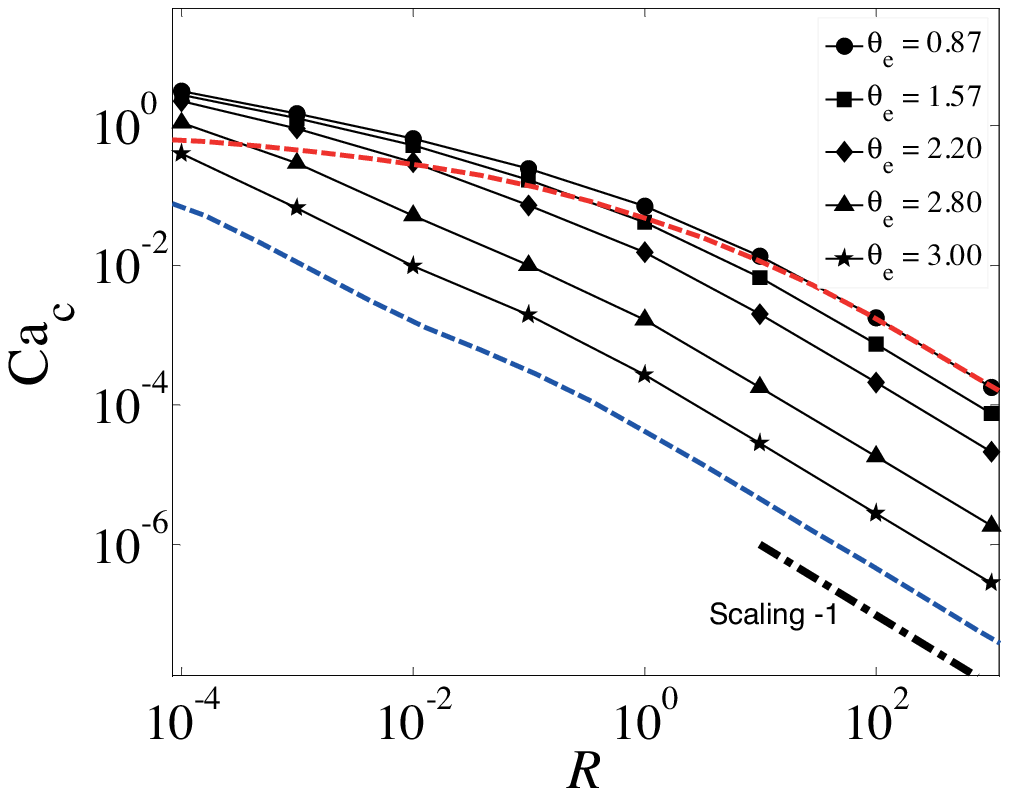}
\caption{\label{cac_R_theta}(Color online) Critical speed $\rm Ca_c$ as a function of
  $R$ for different static contact angle $\theta_e$
  ($\lambda_s=0.001$). Symbols are results of GL model. Dashed curves
  are predictions of Cox's model (top one for $\theta_e=0.87$ radian,
  bottom one for $\theta_e=3.0$ radian), from Eq. (8.3) of \cite{C86},
  for which we take the ratio between the microscopic length scale and
  macroscopic length scale to be $10^{-3}$ (the same value as our slip
  length scaled by the capillary length $\lambda_s$). The
  dashed-dotted line indicates slope of -1.}
\end{center}
\end{figure}

\begin{figure}
\begin{center}
\includegraphics[width=0.65\textwidth]{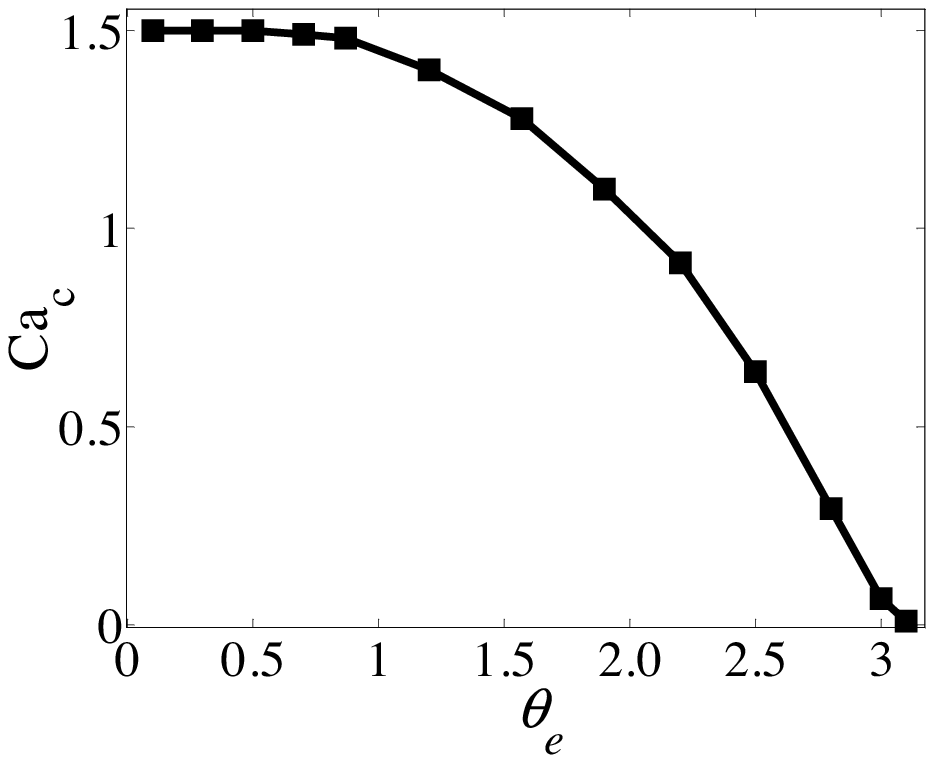}
\caption{\label{cac_thetae}Critical speed $\rm Ca_c$ as a function of
  static contact angle $\theta_e$ for $R=0.001$ and
  $\lambda_s=0.001$.}
\end{center}
\end{figure}

Apart from the viscosity ratio, the GL model contains two parameters:
the slip length $\lambda_s$ and the microscopic (equilibrium) contact
angle $\theta_e$. Here we discuss the dependence of $\rm Ca_c$ on
these parameters. The slip length was varied already in
Fig. \ref{Ca_R}, with values $\lambda_s=10^{-5}$, $10^{-4}$ and
$10^{-3}$.  As expected for wetting problems, we see a weak increase
of $\rm Ca_c$ with $\lambda_s$. A larger $\lambda_s$ reduces the range
over which viscous dissipation is effective. This leads to a
(logarithmic) reduction of the viscous dissipation, while the
capillary driving remains unaltered. Hence, larger velocities can be
achieved before air entrainment occurs.

The dependence of the critical speed on $\theta_e$ is investigated in
Fig. \ref{cac_R_theta}. The figure reveals that there is no obvious
universal scaling behavior for ${\rm Ca}_c$ down to viscosity ratios
as small as $R=10^{-4}$. 
Enforcing a power-law fit, different $\theta_e$ would give rise to
different exponents. We do clearly see that $\rm Ca_c$ decreases with
$\theta_e$, which is further emphasized in
Fig. \ref{cac_thetae}. Consistent with \cite{DYCB07,LRHP11}, the
critical speed vanishes in the hydrophobic limit where $\theta_e
\rightarrow \pi$. We note that for contact angles that are not close
to $\pi$, the shape of the meniscus displays significant
curvatures. In this sense, we expect that our results are not fully
quantitative solutions of the Stokes flow problem.

\section{Discussion and Conclusions}
In this paper we present,
compare and employ two distinct models to study the meniscus deformation and the onset of air entrainment in a dip-coating geometry. The first model is a generalization of the lubrication theory to a two-phase flow situation, in which a slip length is introduced to
resolve the viscous singularity.  The second model is a numerical one
based on the discretization of the Boltzmann equation, namely the
Lattice Boltzmann method for multiphase/multicomponent fluids. In this
model the viscous singularity is removed by the lattice
discretization, which effectively introduces an effective slip length to the system \cite{Srivastava2012}.
The effective slip length also depends on the choice of multiphase/multicomponent model.

The results of GL and LB have a
good agreement, in particular when $\rm Ca$ is relatively small. When
exploring larger values of $\rm Ca$, the two models start to differ as
shown in Fig. \ref{delta_ca_compare}, which can be attributed to
different physics at microscopic and hydrodynamical scales
 (e.g. the non-zero interface thickness in the LB or
   strong viscous bending of interface breaks the
  hypothesis of small interface curvatures made in the GL approach).
Yet, qualitative features, such as the bending of the meniscus and the
dependence on viscosity ratio, are consistent for the two models.
The transition to air entrainment for $\theta_e$ close to $\pi$ involves relatively
weak curvatures and is thus captured by the GL model. 
For the LB simulations the main challenge is given
by the large viscosity contrasts, which is still not fully achievable 
for the multiphase/multicomponent LB model used here.

In the second part of this paper, the critical speed of air
entrainment is investigated by the GL model. We have found a strong
dependence of critical speeds on the air viscosity, which is
consistent with the experiments performed by Marchand \emph{et al.}
\cite{MCSA12}. Remarkably, both our theoretical results and the
experimental results from \cite{MCSA12} differ from Cox's model in
which $\rm Ca_c$ is predicted to depend only logarithmically on air
viscosity \cite{C86}. For comparison, we have added in
Fig. \ref{cac_R_theta} predictions of Cox's model, from Eq. (8.3) of
\cite{C86}, represented by the dashed curves (top curve for
$\theta_e=0.87$ radian, bottom one for $\theta_e=3.0$ radian). For
large $R$, we find that both models predict $\rm Ca_c$ scales as $R$ with
scaling $-1$. This regime corresponds to the usual dewetting case for
which the critical speed only depends on the viscosity of the more viscous
fluid \cite{E04b,SADF07, DFSA07}. Note that in Cox's model, there is an undetermined factor
$\epsilon$ which is defined as the ratio of the microscopic
length scale to the macroscopic length scale. Here we take $\epsilon$ to
be the same value as $\lambda_s$, which is $10^{-3}$. Interestingly,
both the models predict exactly the same values of $\rm Ca_c$ for
$\theta_e=0.87$ radian when $R$ is large, which we consider as an
coincidence since $\epsilon$ is an adjustable parameter. 
For $\theta_e=3.0$ radian, Cox's and our results differ by a
factor. More interesting things occur at small $R$, it is clearly
shown in Fig. \ref{cac_R_theta} that for Cox's model (red top curve )
 $\rm Ca_c$ increases extremely slowly (logarithmically) as
$R$ is decreased. By contrast, our model predicts a moderate increase of $\rm
Ca_c$. Interestingly, such a weak logarithmic relation has been observed in
the case of liquid impacting on liquid \cite{LRQ03,E01}, for which
there is no moving contact line. Both our theoretical results and
experimental results from \cite{MCSA12} therefore suggest that the
mechanism leading to air entrainment can be fundamentally different
depending on whether a contact line is present or not.

The  GL model is directly compared with experiments in \cite{MCSA12},
and shows that  the model is able to  qualitatively capture the dependence of $\rm Ca_c$
on the viscosity ratio $\eta_g/\eta_\ell$. 
Quantitatively, however, the agreement is not satisfactory (see Fig.~3 of \cite{MCSA12}). We
believe this is due to the relatively large meniscus curvatures
encountered in the experiments (static contact angle of the substrate
$\approx 50^o$), pushing the problem beyond the assumptions of the
model.  It would be interesting to explore other methods to achieve a
more quantitative description of air entrainment by advancing contact
line, in particular for large values of $\rm Ca$. From an experimental
perspective, more insight could be obtained by varying the gas
viscosity or by replacing the air by a liquid of low viscosity. It
would also be interesting to perform experiments with a substrate of
large static contact angles.

\section{Acknowledgment}
This work was partially supported by a research program of the
Foundation for Fundamental Research on Matter (FOM), which is part of
the Netherlands Organisation for Scientific Research (NWO). 
T.S.C. acknowledges financial support by the FP7 Marie Curie Initial Training 
Network Surface Physics for Advanced Manufacturing project ITN 215723. We thank useful
discussions with M. Sbragaglia. L.Biferale acknowledges support from
DROEMU-FP7 IDEAS Contract No. 279004. B. Andreotti is supported by Institut
Universitaire de France.

\appendix
\section{Appendix}
We discuss the behavior of the function $f(\theta,R)$ when $\theta$ is close to $\pi$ and $R$ close to zero. As presented in Section \ref{LTmodel_derivation}, $f(\theta,R)$ is defined as:

\begin{eqnarray}
f(\theta,R)&\equiv& \frac{2\sin^{3}\theta[R^{2}f_1(\theta)+2R
    f_3(\theta)+f_1(\pi-\theta)]}{
  3[Rf_1(\theta)f_2(\pi-\theta)-f_1(\pi-\theta)f_2(\theta)]}
\nonumber\\ f_1(\theta)&\equiv&\theta^{2}-\sin^{2}
\theta\nonumber\\ f_2(\theta)&\equiv&\theta-\sin\theta
\cos\theta\nonumber\\ f_3(\theta)&\equiv&(\theta(\pi-
\theta)+\sin^{2}\theta).\nonumber\\
\end{eqnarray}


First, we expand the terms in both the numerator and the denominator in series of ($\pi-\theta$) and keep the leading order terms only, we end up with
\begin{equation}
f(\theta,R)\simeq\frac{-2[\pi^2R^2+2\pi R(\pi-\theta)+(\pi-\theta)^4/3]}{2\pi^2R+\pi(\pi-\theta)}.
\end{equation}

The asymptotic behavior of $f(\theta,R)$ depends on the relative magnitude between $(\pi-\theta)$ and $R$.
For $R\ll\pi-\theta$, $f(\theta,R)$ can be approximated as: 
\begin{equation}
f(\theta,R)\simeq f(\theta,0)-4R\simeq\frac{-2(\pi-\theta)^3}{3\pi}-4R.
\end{equation}
The contribution of air viscosity, represented by $-4R$, will become significant once $(\pi-\theta)\sim R^{1/3}$.

When $\theta$ is very close to $\pi$ such that $\pi-\theta\ll R$, $f(\theta,R)$ goes to a different asymptotic form, i.e.
\begin{equation}
f(\theta,R)\simeq -R.
\end{equation}
If we substitute this asymptotic form of $f(\theta,R)$ into the generalized lubrication equation (\ref{full_interface_eqn}), we will see the liquid viscosity will be canceled out in the multiple $\rm Ca R$ so that the asymptotic equation does not depend on liquid viscosity. This means in this asymptotic limit, air viscosity completely dominates the flow.

\bibliographystyle{unsrt}
\bibliography{all_ref}
\end{document}